\documentclass{aastex}

\shorttitle{Polarization Patterns in CMBR}
\shortauthors{Chiueh}

\begin{document}

\title{On the Local Separation of $E$ and $B$ Polarization Patterns in CMBR}


\author{Tzihong Chiueh}

\affil{Physics Department, National Taiwan University, Taipei,
Taiwan} 
\affil{Institute of Astronomy and Astrophysics, Academia
Sinica, Taipei, Taiwan} 
\email{chiuehth@phys.ntu.edu.tw}


\begin{abstract}

This work examines the small-scale $E$ and $B$ patterns of the Cosmic Microwave 
Background Radiation (CMBR)
polarization anisotropy.  Particularly, we address the topological natures
of the $E$ and $B$ modes, and how one may make use of the
local measurements of Stokes parameters to separate these two modes 
in the real space.  The analysis of a local map in the Fourier space for 
separating the $E$ and $B$ modes can be an
ill-posed problem, due to the non-periodic boundary condition in the map for 
each mode.  A strategy for the $E-B$ separation in a local map is through an 
appropriate
projection of the polarization tensor field into a unique vector field, which
naturally contains a curl-free $E$ component and a divergence-free $B$ 
component.  An integral method, equivalent to the real-space top-hat filtering,  
is proposed for extracting the averaged "charge" and "current" of
the $E$ and $B$ modes separately.  The real-space top-hat filter function exhibits, 
in its power spectrum, an oscillation of a well-defined frequency comparable to
the filter size.  It is pointed out that when the inverse filter size is chosen
properly, matching the oscillation period in the $E$-mode power spectrum,
the sensitivity of the $E$-mode detection can be made to
be significantly improved.

\end{abstract}


\keywords{Cosmic Microwave Background --- Polarization}


\section{Introduction}

Anisotropy of the CMBR is linearly polarized\citep{ree68,bon84,pol85}.
The linear polarization was produced by the Thomson
scattering of the local temperature quadrupole anisotropy off the
uniform background electrons at the last scattering surface, where the optical
depth was about $1/2$.  The angular pattern of the CMBR linear polarization
therefore contains the imprints of the temperature anisotropy.  However,
unlike the temperature anisotropy, which is a scalar, the polarization is a
tensor. The angular pattern of tensor fluctuations can generally be decomposed
into an electric-type pattern ($E$-mode) and a magnetic-type pattern
($B$-mode), according to their distinct global parity
symmetries\citep{kam97,zal97}.

Despite a handful of past works that have been devoted to the
detailed analyses of CMBR polarization for the $E$ and
$B$ modes in the Fourier space\citep{kam97,zal97,hu97a,hu97b},
few of them gave clear descriptions of what these unfamiliar $E$
and $B$ polarization patterns really are in the real space.  
These works followed the standard paradigm
that the initial fluctuations are Gaussian with no feature, and hence the
polarization Fourier spectra were the sole relevant quantities of
investigations.   Hu and White (1997b) gave, by far, the most intuitive picture
of what these two types of polarization patterns are on the large scale,
but their work on the subject of how the $E$ and $B$ modes 
may be separated is still confined to the power-spectrum method 
and still under the assumption of Gaussian fluctuations.

In the coming dedicated experiments for CMBR polarization \citep{sta99,lo00},
deep maps of finite patches of local
sky are to be constructed.  Though the $E$ and $B$ modes have conventionally
be defined in the spin harmonics space\citep{gol67} or the Fourier space, 
it is in fact not
desirable to analyze a map of finite sky coverage in such a space.
This is because, unlike the global map, the local map does not satisfy the
periodic boundary condition and the image has to be distorted in order for
the Fourier transformation to be carried out, thereby making the detection of 
the predicted small $B$ mode to be unreliable.   Merely for this reason,
it has already had the real need to call for the real-space analysis of the CMBR polarization maps.

The real-space analysis also has the advantage of detecting localized
non-Gaussian features, which may lead to discovery of the cosmic objects
not predicted by the standard inflationary cosmology, such as the cosmic 
strings \citep{sel97}.  Even in the framework of the standard
cold-dark-matter paradigm, where the CMBR polarization is produced almost
instantaneously at the last scattering surface, the polarized sky should
contain a clean signature of the  coherent patches of Hubble
spheres at the epoch of photon-electron decoupling. 
This expectation is signified by the large-amplitude oscillation of a well-defined 
period in the
polarization power spectrum.  The real-space analysis can thus be potentially 
better methodology
than the Fourier (l) space analysis to capture a large fraction of the total 
power within the
$l$-space oscillation by using some properly tailored real-space filter
functions, a subject that will be addressed in the last section.

Despite these advantages,
the direct means for measuring the $E$ and $B$ modes in the
real-space maps has not been available.  
It is in this context that the CMBR polarization in the real space is
addressed in this report.  To begin, I shall briefly review the CMBR
polarization, followed by a more rigorous description of the $E$ and $B$ modes
in the real space.  After that, a direct means for disentangling these two
modes is proposed.  

\section{Spatial Symmetry}

Unlike the brightness, which is a scalar, the linear polarization 
cannot distribute itself uniformly on the celestial sphere, due to the
curvature of a sphere.  In other words, one can never arrange line segments
to be uniformly distributed on a spherical surface.   This statement echoes the
more familiar notation that never can two families
of lines lay out a globally {\it uniform} coordinate on a spherical surface; the
best one can do is the longitude-latitude coordinate, which has quadrupole
inhomogeneity.  For vector fields, the highest spatial symmetry
has a dipole pattern, and for the polarization, a tensor field, the highest symmetry
has a quadrupole pattern.  Therefore the CMBR
linear polarization must be angular dependent and must be a field with angular
patterns, at scales smaller than or equal to the quadrupole, on the sky.  

Though the instantaneous electric field is indeed a vector, the measured electric 
field of a polarized light is not a
vector, since the measured electric field can be in either direction, a
"bi-vector".  With an 180-degree rotation of the "bi-vector",
the polarization can recover itself but a true vector can not.  Such a 
peculiar bi-vector can be constructed from a second-rank polarization tensor,
as shown below. 
Similar to a two-dimensional vector, the polarization tensor can also be
described by two parameters --- the magnitude and orientation of the bi-vector.
To distinguish a vector ${\bf V}$ from a polarization tensor ${\bf P}$,
the former is usually expressed as a two-component array, or 
${\bf V}=V_x\hat x+V_y\hat y$, and the latter as a traceless
$2\times 2$ matrix, or  ${\bf P}=P_x\sigma_1 + P_y\sigma_2$, where
$\sigma_1$ and $\sigma_2$ are respectively the $z$ and $x$ components of the
Pauli matrices. (Assume the line of sight to be along the $z$ direction.) 
The components $P_x$ and $P_y$ are the Stokes parameters, $Q$ and $U$, 
respectively, and the ratio $U/Q$ is related to the orientation of polarization.
The angle between the polarized electric field and the $x$ axis is
$\theta=(1/2)\tan^{-1}(U/Q)$, which is bi-directional and has an $180$-degree
directional ambiguity, and $(Q^2+U^2)^{1/4}$ is the amplitude 
of the polarized electric field.  

As the polarization field ${\bf P}(z,x)$ must have spatial patterns on the
sky, much like the more familiar vector field,
the polarization field can be decomposed into two
coordinate-independent components of different topologies.  For the
small-scale angular pattern, one may adopt the sky-flat approximation. 
The two topologically distinct components of a two-dimensional
{\it vector} field are the curl-free $\nabla\phi(x,y)$ and 
divergence-free $\nabla\times(\psi(x,y)\hat z)$
vector fields.  However, the decomposition for the polarization field is not
obvious, and it needs to be guided by some
principles of spatial symmetry.  

We shall first examine the familiar vector field 
in an attempt to extracting some guiding principles.  Consider first the 
parity symmetry. The spatial inversion $(x,y)\to(-x,-y)$ yields
$\nabla\phi\to -\nabla\phi$ and $\nabla\times(\psi\hat z)\to
-\nabla\times(\psi\hat z)$, and hence such an operation can not distinguish
the two components.  In fact, the visual distinction of a diverging pattern
from a swirling pattern arises from partial spatial inversion, $x\to -x$ or 
$y\to -y$.   The $x$ and $y$ components of a diverging vector field transform in
accordance with how the two coordinates change, but those of a swirling vector
field transform just oppositely; that is, flipping the sign of the $x$ coordinate 
changes only the sign of the $y$ component of the swirling vector and flipping
the sign of the $y$ coordinate changes only the sign of the $x$ component.  
The partial space symmetry clearly shows the difference of
a two-dimensional vector (the gradient) from a two-dimensional pseudo-vector (the curl).

A traceless second-rank tensor can be regarded as a bilinear combination of two
vectors, and hence it is possible to construct a vector from a tensor, where
the constructed vector contains the topological features of the original tensor.
To see how this is possible, we now consider the polarization tensor specifically.

\section{Topological Characteristics in the Polarization Tensor Field} 

Let the measured electric field be
\begin{equation}
\langle E_i({\bf x})E_j({\bf x})\rangle=I_0({\bf x})\delta_{ij}+Q({\bf
x})\sigma_1+ U({\bf x})\sigma_2,
\end{equation}
where $I_0$ describes the anisotropy intensity and the traceless polarization 
tensor ${\bf P}$ is described by the last two terms on the right.
We may also express the measured electric field equally well as
\begin{equation}
\langle E_i({\bf x})E_j({\bf x})\rangle=I_0'({\bf x})\delta_{ij}+2p_i({\bf
x})p_j({\bf x}),
\end{equation}
where ${\bf p}({\bf x})$ is a two-dimensional vector field.
Similar to Eq.(1), $\langle E_i E_j\rangle$ in
Eq.(2) is also characterized by three independent parameters for a general
$2\times 2$ symmetric real tensor, i.e., the unpolarized intensity
$I_0'=I_0-(p_x^2+p_y^2)$, and the linearly polarized
components $Q=p_x^2-p_y^2$ and $U=2p_xp_y$. 

Though such a decomposition of a tensor field ${\bf P}$ into a bilinear combination
of a vector ${\bf p}$ is legitimate, ${\bf p}$ is not
unique since $-{\bf p}$ can be an equally good choice, and this reflects
the bi-directional nature of the polarization of ${\bf p}$.  
It is therefore of crucial
importance to determine whether the polarization tensor ${\bf P}$ is the primary
field which can mathematically be decomposed into the outer product
of two vectors, or the vector field ${\bf p}$ is the primary field which,
with a bilinear combination, can form a second-rank tensor.   
For the latter, the vector ${\bf p}$ and the tensor ${\bf P}$ are both 
well-defined,
but for the former, ${\bf p}$ is an ambiguous bi-vector.   In the
CMBR, it is generally believed that the polarization tensor, i.e., the
Stokes $Q$ and $U$, is the primary field produced by the temperature
quadrupole anisotropy.  That is, the bi-vector ${\bf p}$ is secondary, constructed 
from the primary tensor ${\bf P}$.  
Whether the ${\bf p}$ field is secondary or not 
can in fact be tested by examining whether there exist 
singularities in ${\bf p}$\citep{nas98}.  This is because the primary tensor 
field ${\bf P}$ vanishes locally
most likely in a linear manner, i.e., non-vanishing first derivatives at the nulls;
it therefore demands ${\bf p}$ to vanish non-analytically
due to the bilinear combination of ${\bf p}$, thus resulting in the branch points in 
${\bf p}$.

With this understanding of the CMB polarization field ${\bf P}$, it becomes
clear that using the vector ${\bf p}$ to describe ${\bf P}$, as written in
Eq.(2), can not be appropriate, and it calls for a suitable representation of the
second-rank ${\bf P}$ tensor field.   To avoid any singularity,
this representation should be linear with ${\bf P}$, and the only possibility to 
construct a tensor via linear operations is through the spatial derivatives
that act as vectors.  One may construct such a second-rank tensor
either by two spatial derivatives acting on a scalar field, or by one
spatial derivative acting on a vector field.   In two-dimensions, a vector
field can always be represented by one spatial derivative on the
scalar fields as shown above, and the two constructions are therefore equivalent. 
From now on, we adopt the former.   The most general representation of
a traceless tensor field of this kind in a flat two-dimensional space can 
be written as:
\begin{equation} 
{\bf P}(x,y)=[\nabla\nabla-(\hat z\times\nabla)(\hat z\times\nabla)] f(x,y)
+[(\hat z\times\nabla)\nabla+\nabla(\hat z\times\nabla)]g(x,y).   
\end{equation}
where $f$ and $g$ are two independent scalar-field components.   Much like the
two independent topological patterns in a vector field, these two components
in the tensor field ${\bf P}$ also have their own topological characteristics;
the former is called the $E$ mode and the latter the $B$ mode.  The different
spatial partial symmetries are revealed in the difference in $\nabla$ and
$\hat z\times\nabla$.  Having this expression of ${\bf P}$, it becomes possible 
to construct a two-dimensional vector from the tensor ${\bf P}$ by contracting 
it with a vector 
differential operator, either with a divergence operator, a curl operator or a combination of both.
However, not all vector differential operators are able to capture the topological
features of the tensor;
only a unique choice of vector differential operator is.

A convenient and concise way to reveal such a choice of vector
differential operator is in the complex representation, where
\begin{equation}
{\bf P}(x,y)=Q(x,y)+iU(x,y)=\frac{\partial^2}{\partial\bar z^2}
[f(x,y)+ig(x,y)],  
\end{equation}
with $z\equiv x+iy$ and $\bar z\equiv x-iy$.
This representation agrees with
the $E$ (associated with $f$) and $B$ (associated with $g$) modes defined above and those defined in
the Fourier space \citep{whi99}.  It also shows clearly how  
$\partial/\partial\bar z$ may act onto ${\bf P}$ to form a vector field ${\bf V}$:
\begin{equation}
{\bf V}\equiv\frac{\partial}{\partial z}{\bf P}=\frac{\partial}{\partial\bar z}
[\nabla^2(f+ig)]=(\frac{\partial Q}{\partial x}+\frac{\partial U}{\partial y})
+i(\frac{\partial U}{\partial x}-\frac{\partial Q}{\partial y}),
\end{equation}
where $\partial^2/\partial z\partial\bar z=\nabla^2$ has been used and
the last equality is given by the definition ${\bf P}\equiv Q+iU$.
The first equality shows clearly that when $g=0$, the resulting ${\bf V}$ 
is a diverging
vector, and when $f=0$, ${\bf V}$ a swirling vector.  Thus, the two topologically
different characteristics of a vector field is also contained in the polarization
tensor field through the components of $f$ and $g$.

\section{Separation of $E$ and $B$ Modes}

In conventional vector notation, Eq.(5) is re-written as
\begin{equation}
{\bf V}(x,y)=
\hat x(\frac{\partial Q(x,y)}{\partial x}+\frac{\partial U(x,y)}{\partial y})
+\hat y(\frac{\partial U(x,y)}{\partial x}-\frac{\partial Q(x,y)}{\partial y})
=\nabla\phi(x,y)+\hat z\times\nabla\psi(x,y),
\end{equation}
where $\phi\equiv\nabla^2 f$ and $\psi\equiv \nabla^2 g$. 
The real-space separation of the $E$ and $B$ modes may be conducted by employing
suitable closed-loop integrations.  Namely, using the Stokes theorem for a 
closed-line integral,  
\begin{equation}
\hat z\cdot\int {\bf V}(x,y)\times d{\bf l} = \int \nabla^2\phi(x,y) d^2S,
\end{equation}
it projects the $E$-mode "charge" (or $\nabla^2\phi$) averaged over a 
top-hat filter enclosed by the loop.  On the other hand, the closed-line integral
\begin{equation}
\int {\bf V}(x,y)\cdot d{\bf l} = \int \nabla^2\psi(x,y) d^2S
\end{equation}
projects the top-hat-filter averaged $B$-mode "current" (or $\nabla^2\psi$).
The variances of the top-hat-filter averaged "charge" and "current" for various
filter sizes can serve as alternative representations in place of 
the power spectra of $E$ 
and $B$ modes.  This representation has an advantage
over the Fourier spectral representation and will be discussed in the
next section.

It is worth noting that though the construction of ${\bf V}$ from the
polarization tensor ${\bf P}$ involves one spatial derivative, which 
amplifies the measurement shot noises, the proposed projection of either mode
involves one spatial integration.  The two operations formally compensate, 
though in actual operations some small level of noises can be
introduced.

\section{Discussions}

In this work, we show the topological natures of the $E$ and $B$ modes by
suitably projecting the polarization tensor into a vector.  The $E$ and $B$
modes correspond to the diverging and swirling patterns of this vector,
respectively.  Having them identified in the real space, an integral method
then becomes a natural choice for separating these two modes, as such a 
method can minimize amplification of the shot noises inherent in the
CMBR measurements.  

Moreover, this real-space analysis of CMBR polarization can retain
the characteristics of $E$ and $B$ even when the measurements are subject
to the detector beam smearing, giving rise to the convolution of the
sky image with the detector beam pattern.  The measured quantities are
the Stokes $Q$ and $U$, and when they are subject to identical beam
smearing, they become $\langle Q\rangle$ and $\langle U\rangle$, where
\begin{equation} 
\langle A\rangle({\bf x}) \equiv \int W({\bf x}+{\bf r}) A({\bf r}) d^2 r
\end{equation}
with $W$ being the beam pattern.   It is easy to see that if $A({\bf r}) =
\partial a({\bf r})/\partial {\bf r}$, the smeared quantity 
$\langle A\rangle({\bf x})=\partial\langle a\rangle/\partial{\bf x}$,
which retains the same differential form, and it does so even when 
higher derivatives are involved.    
That is, the tensor decribed by Eq.(4) can
retain the same topological nature with beam smearing.  Moreover since
Eq.(4) is linear in $f$ and $g$ and the convolution is a linear
operation,  it follows that the $E$ and
$B$ modes can not be mixed by beam smearing.   

The pixelization of image 
is another issue in the real-space analysis, and it involves a particular
form of the window function  $W({\bf x},{\bf r})=\sum_i  \delta({\bf x}-{\bf
x}_i) W_0({\bf x}+{\bf r})$.   The Stokes $\langle Q\rangle$ and 
$\langle U\rangle$ can also maintain their two components of distinct
topologies unmixed.  However, further manipulations with differentiation,
approximated by differencing, and integration, approximated by summation,
should proceed with caution to keep $E$ and $B$ intact.  We will leave
this technical issue to a future paper.

As mentioned in the last section, 
the projection with a closed-loop line integral corresponds to a top-hat-filter
average of the polarization signal.  The Fourier spectrum of this filter function
is proportional to $J_1^2(kR)$, which 
oscillates in the $k$ space with a frequency comparable to the filter size
$R$, where $J_1$ is the Bessel function of order one and $k$ the Fourier 
wavenumber.
It is noted that in the CDM cosmology, with or without the cosmological
constant, the power spectrum of the $E$ polarization also oscillates with a
well-defined period in the $k$-space.  It is therefore anticipated that by
varying the radius $R$ of a circular integration loop of Eq.(7), the
frequency of the filter power spectrum can be made to match that of  the
$E$-mode power spectrum, where the response to polarization signals is
maximal.   Such an analysis strategy has not been considered, or even
mentioned, in the past,  mainly because most attention has so far been
directed toward the Fourier-space analysis of CMBR signals.  

The Fourier-space analysis can be good methodology for analyzing
the CMBR temperature anisotropy, since the acoustic oscillations in the
spectrum have an ill-defined $k$-space frequency, due to the changing
Hubble radius when the acoustic waves entered the horizon.  However, the
polarization in the CMBR results from a much cleaner physics than the
temperature anisotropy does.  The polarization
involves only the Thomson scattering of anisotropic photons at the epoch when
the mean-free-path was about the then Hubble radius.  Any structure
of size smaller than the then Hubble radius is smeared out by the large Thomson
mean-free-path, and any structure of size greater than the then Hubble radius
has a negligible power.   This characteristic Hubble sphere enters the phase
of the perturbation, and naturally gives rise an oscillation
in the power spectrum with a well-defined frequency comparable to the then
Hubble radius.  It is therefore of little surprise that a real-space top-hat
filter function can capture the Hubble spheres and be well suited for
detecting the $E$ polarization signals.  A detailed analysis of how the
presently proposed method performs for detecting the CMBR polarization will  be
reported elsewhere.  
\acknowledgments
 
I would like to thank Cheng-Jiun Ma for helpful discussions.
This work is supported in part by the National Science Council of Taiwan
under the grant NSC89-M-2112-002-065.




\end{document}